\documentclass[runningheads]{llncs}
\usepackage[T1]{fontenc}
\usepackage{graphicx}
\usepackage{caption}
\usepackage[table]{xcolor}
\usepackage{array}
\usepackage{graphicx} 
\usepackage{float}
\usepackage{amsmath,amssymb,bm}
\usepackage{hyperref}
\usepackage{url}
\usepackage{multirow}
\usepackage{booktabs}
\usepackage[final]{microtype}
\usepackage{enumitem} 
\usepackage{hyperref} 
\newcommand{\myuparrow}{\ensuremath{^{\textcolor{red}{\textbf{\text{↑↑}}}}}}
\newcommand{\mydownarrow}{\ensuremath{^{\textcolor{blue}{\textbf{\text{↓↓}}}}}}
\usepackage{hyperref}
\begin{document}

\title{Segmenting Thalamic Nuclei:\\ $\mathrm{T_1}$ Maps Provide a Reliable and Efficient Solution}

\author{
Anqi Feng\inst{1, 2} \and
Zhangxing Bian \inst{1} \and
Samuel W. Remedios \inst{3} \and 
Savannah P. Hays \inst{1} \and 
Blake E. Dewey \inst{4} \and 
Jiachen Zhuo \inst{5} \and 
Dan Benjamini \inst{2} \and
Jerry L. Prince \inst{1}
}
\institute{
Department of Electrical and Computer Engineering, \\Johns Hopkins University, Baltimore, USA\\[0.4em] \and
Laboratory of Behavioral Neuroscience, National Institute on Aging, National~Institutes~of~Health, Baltimore, USA\\[0.4em] \and
Department of Computer Science, Johns Hopkins University, Baltimore, USA \and
Department of Neurology, Johns Hopkins School of Medicine, Baltimore,~USA\\[0.4em] \and
Department of Diagnostic Radiology and Nuclear Medicine,
University~of~Maryland~School~of~Medicine, Baltimore, USA
}

\authorrunning{A. Feng et al.}
\titlerunning{Segmenting Thalamic Nuclei with $\mathrm{T_1}$ Maps}
\maketitle  
\begin{abstract}
{Accurate thalamic nuclei segmentation is crucial for understanding neurological diseases, brain functions, and guiding clinical interventions. However, the optimal inputs for segmentation remain unclear. This study systematically evaluates multiple MRI contrasts, including MPRAGE and FGATIR sequences, quantitative PD and $\mathrm{T_1}$ maps, and multiple T1-weighted images at different inversion times (multi-TI), to determine the most effective inputs. For multi-TI images, we employ a gradient-based saliency analysis with Monte Carlo dropout and propose an Overall Importance Score to select the images contributing most to segmentation. A 3D U-Net is trained on each of these configurations. Results show that $\mathrm{T_1}$ maps alone achieve strong quantitative performance and superior qualitative outcomes, while PD maps offer no added value.
These findings underscore the value of $\mathrm{T_1}$ maps as a reliable and efficient input among the evaluated options, providing valuable guidance for optimizing imaging protocols when thalamic structures are of clinical or research interest. Codes are available at \href{https://github.com/ANQIFENG/T1map-for-Thalamus}{T1map-for-Thalamus}.
}
\end{abstract}

\section{Introduction}
The thalamus is a deep brain structure consisting of multiple distinct nuclei, each specializing in functions and involved in neurological disorders and treatments~\cite{sherman2006exploring,sherman2001exploring}. For example, the ventral lateral posterior nucleus regulates motor control~\cite{rispal1973relations,sommer2003role}; 
the limbic nuclei relate to Alzheimer’s disease~\cite{aggleton2016thalamic,braak1991alzheimer}; and the ventral intermediate nucleus  is a key target for deep brain simulation (DBS) in essential tremor~\cite{kumar2003long,papavassiliou2008thalamic}. Given these roles, accurate segmentation of thalamic nuclei is important for both clinical interventions and neurological research.

\begin{figure}[!tb]
\centering
\includegraphics[width =0.95 \textwidth]{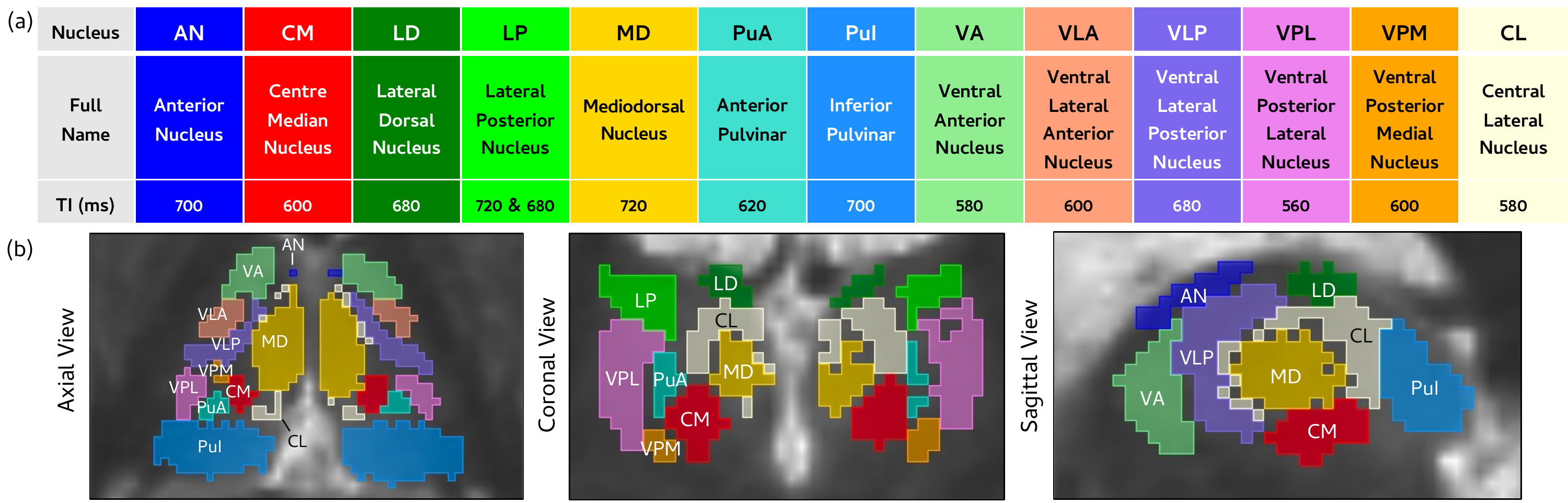}
\caption{(a) Thalamic nuclei with their color codes, and corresponding inversion times (TI) used for manual delineation. (b) Sparse labels overlaid on the $\mathrm{T_1}$ map.}
\label{fig:s2-label_naming_and_maps_used}
\end{figure}

In-vivo thalamic nuclei segmentation primarily relies on Magnetic Resonance Imaging (MRI). 
However, low intra-thalamic contrast and small nuclei size make accurate segmentation challenging with standard T1-weighted (T1w) and T2-weighted (T2w) sequences. 
Magnetization-Prepared Rapid Gradient Echo (MPRAGE)~\cite{mugler1990three} has been widely used but provides limited contrast within the thalamus and against surrounding white matter and subcortical structures~\cite{iglesias2018probabilistic,umapathy2022convolutional}.
Fast Gray Matter Acquisition T1 Inversion Recovery (FGATIR)~\cite{sudhyadhom2009high} has been shown to enhance intra-thalamic contrast by nulling white matter~\cite{saranathan2021vivo,su2019thalamus,tourdias2014visualization}, though the shorter inversion time can lead to a somewhat lower signal-to-noise (SNR) ratio compared to MPRAGE~\cite{tourdias2014visualization}.
Quantitative
MRI, which can yield intrinsic tissue properties such as $\mathrm{T_1}$, $\mathrm{T_2}$, and proton density (PD) maps, has also been explored~\cite{deoni2007segmentation,traynor2011segmentation}. $\mathrm{T_1}$ maps, in particular, reveal relaxation differences between nuclei~\cite{deoni2005visualization,tourdias2014visualization}, facilitating visual discrimination of major nuclear groups.
Some methods acquire multiple T1-weighted images at different inversion times (multi-TI), each tailored to null specific tissue compartments at different timepoints, capturing finer contrast variations among nuclei~\cite{feng2024ratnus,feng2023label,yan2023segmenting}.
Diffusion MRI (dMRI) has been widely used in~\cite{duan2007thalamus,najdenovska2018vivo,wiegell2003automatic}, but the low anisotropy in thalamic gray matter and the limited spatial resolution of Echo Planar Imaging (EPI) hinder accurate nuclei localization.
Connectivity-based segmentation is clinically useful for DBS targeting~\cite{johansen2005functional,o2011clustering}, but often misaligns with anatomical boundaries. Many adjacent nuclei project to similar cortical targets, grouping distinct nuclei together, while individual nuclei with diverse projection patterns may split into multiple clusters.

Multi-TI images offer diverse tissue contrasts across inversion times. However, not all are equally informative for thalamic nuclei segmentation. Selecting the most relevant images is necessary, but training a separate model for every possible combination is impractical. Gradient-based methods are widely used to generate post-hoc attribution maps for identifying important image regions~\cite{selvaraju2017grad,sundararajan2017axiomatic}, and have been extended to assess the importance of multimodal medical images~\cite{kawahara2017brainnetcnn}. Monte Carlo Dropout~\cite{gal2016dropout} estimates prediction uncertainty at test time through multiple stochastic forward passes and has been applied in medical imaging tasks~\cite{eaton2018towards,nair2020exploring}. However, integrating gradient-based saliency analysis with Monte Carlo Dropout to quantify image importance remains underexplored.

Despite previous efforts, the optimal MRI contrast for thalamic nuclei segmentation is still unclear. In this study, we focus on T1w-based images, systematically comparing structural sequences (MPRAGE, FGATIR), quantitative maps (PD, $\mathrm{T_1}$ maps), and multi-TI images. For multi-TI images, we perform gradient-based saliency analysis across multiple Monte Carlo dropout runs and define an Overall Importance Score to select the most informative inversion times for segmentation.
To the best of our knowledge, this is the first study to leverage gradient-based saliency analysis with Monte Carlo Dropout to guide input feature selection. Our results show that the $\mathrm{T_1}$ maps alone strike a favorable balance between accuracy and efficiency, making it a practical choice for thalamic nuclei segmentation.

\section{Methods}
\subsection{Data and Manual Labels}
Our MRI data were collected as part of a mild traumatic brain injury (MTBI) study, approved by the local ethics board. It involves 24 participants: 14 healthy controls and 10 individuals with MTBI. Each participant underwent an MRI session with MPRAGE and FGATIR, both acquired at 1 mm isotropic resolution, repetition time (TR) 4000 ms, and echo time (TE) 3.37 ms. The only difference was in the inversion time (TI): 1400 ms for MPRAGE and 400 ms for FGATIR.

Manual thalamic nuclei delineation was guided by the Morel Atlas~\cite{morel1997multiarchitectonic}, focusing on 13 nuclei (or nuclear groups). Multi-TI images were used for annotation due to their enhanced contrast for these small, densely packed structures. For each nucleus, the rater selected the most visually distinct TI sequence to maximize delineation clarity, then verified alignment with $\mathrm{T_1}$ maps. Figure~\ref{fig:s2-label_naming_and_maps_used}(a) lists the nuclei names along with the chosen TI values. Only voxels with high-confidence label assignments were annotated, resulting in relatively sparse labels and leaving many unlabeled voxels near the thalamic boundaries (Figure~\ref{fig:s2-label_naming_and_maps_used}(b)).

\subsection{Data Processing and Computation}
We preprocess MPRAGE and FGATIR following the procedure in~\cite{feng2024ratnus} to generate PD and $\mathrm{T_1}$ maps, and multi-TI images. MPRAGE and FGATIR are co-registered to MNI space using ANTS~\cite{avants2009advanced}. Both images then undergo N4 Bias Field Correction and White Matter Mean Normalization. Notably, these images are processed jointly using a harmonic bias field and a common scaling factor determined from a white matter mask derived from MPRAGE, to ensure accurate maps estimation.

\begin{figure}[!tb]
\centering
\includegraphics[width = 0.9\textwidth]{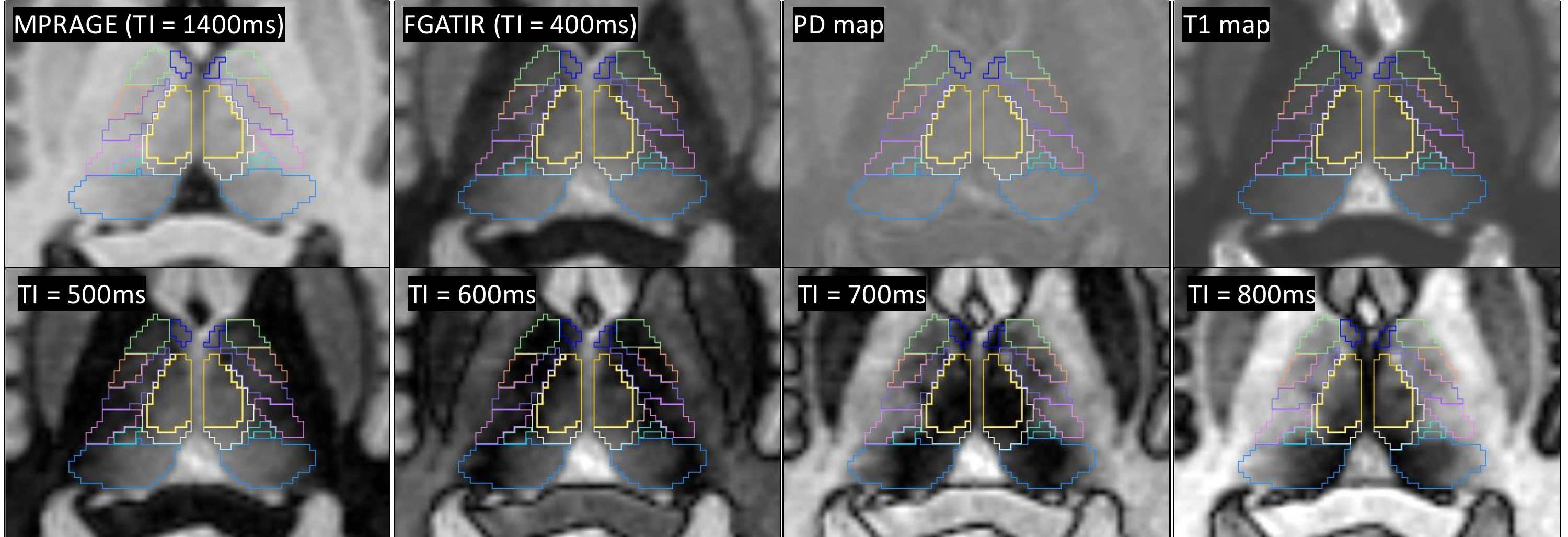}
\caption{MPRAGE, FGATIR, PD and $\mathrm{T_1}$ maps, and example multi-TI images.}
\label{fig:s2-imaging_data_visualization}
\end{figure}

In a T1‐weighted MRI, a voxel $v$’s intensity $I{(v)}$ can be modeled as:
\begin{equation}
I{(v)} = \mathrm{PD}{(v)} [ 1 - 2e^{(-\mathrm{TI} / T_1{(v)})} + e^{(-\mathrm{TR} / T_1{(v)})} ], 
\label{formula_1}
\end{equation}
where $\mathrm{PD}{(v)}$ is the voxel’s proton density, $\mathrm{T_1}{(v)}$ the T1 relaxation time, TI the inversion time, and TR the repetition time. As described in~\cite{feng2024ratnus}, by acquiring MPRAGE and FGATIR with  the same parameters but different TIs, we can determine each voxel’s $\mathrm{PD}$ and $\mathrm{T_1}$. The $\mathrm{PD}$ and $\mathrm{T_1}$ maps enable further computation of T1‐weighted images at any chosen TI. In Figure~\ref{fig:s2-imaging_data_visualization}, example multi-TI images show how  TI variations affect tissue contrast and highlight different features.

\subsection{Identify the Optimal Inputs}
\label{subsec:Identify the Optimal Inputs}
We aim to identify the most effective input modalities for thalamic nuclei segmentation through a two-stage process. Stage 1 selects the most informative multi-TI images. Stage 2 compares structural sequences (MPRAGE, FGATIR) against quantitative maps (PD, $\mathrm{T_1}$ maps) and the selected multi-TI images.\\
\textbf{Stage 1: Selecting informative multi-TI images} \\ 
\underline{(1) Segmentation Model and Input Images:}
A 3D U-Net~\cite{cciccek20163d} with Dice loss is adopted to segment the thalamus into 14 classes, including 13 nuclei (or nuclear groups) and background. 
We trained the model on 53 3D images per subject. These include: 
1) 51 multi-TI images, computed at TIs from 400ms to 1400ms in 20ms increments. This range captures meaningful contrast differences in thalamic structures, with 20ms as the smallest visually discernible step. 
2) MPRAGE and FGATIR, corresponding to the 1400 ms and 400 ms TIs, respectively. 
While the multi-TI images are derived from quantitative maps, MPRAGE and FGATIR are directly acquired. All 53 images are treated as distinct input contrasts.\\
\underline{(2) Gradient Analysis with Monte Carlo Dropout:}
To quantify the contribution of each input image, we employ gradient-based saliency analysis during inference. 
Specifically, we compute the gradients of the model’s output probability for each class with respect to input images. 
Unlike traditional pixel-wise saliency maps that focus on individual voxel contributions, our approach captures channel-wise importance. 
Larger gradients indicate stronger influence on the model’s prediction, while smaller gradients reflect weaker contributions. 
This method is interpretable and computationally efficient, as it leverages the network’s existing backpropagation mechanism to reveal each input’s relevance without adding extra computational overhead.
However, single gradient computations can be sensitive to factors such as model architecture, random initialization, and inference-time stochasticity. 
These sources of variability may lead to unstable estimates of input importance.
To address this, we apply Monte Carlo (MC) Dropout during inference, performing multiple forward-backward passes with dropout enabled. 
This simulates an ensemble of different “sub-models” without training separate networks, efficiently capturing variability across runs. 
Gradients computed across multiple backward passes are then integrated to obtain robust estimates of each image’s contribution, guiding the selection of the most informative inputs.\\
\underline{(3) Overall Importance Score (OIS):}
We define the OIS to summarize the gradient information and rank the image contributions.
\[
\text{OIS}_i = \frac{1}{SM} \sum_{(c, s, m) \in \mathcal{X}} \left| \nabla_{I_{i,s}} \sum_{v \in \mathcal{V}_s} f_\theta^{(m)}(I_{\cdot,s})[c, v] \right|, \quad \text{where } \mathcal{X} = C \times S \times M
\]

\noindent
where:
\begin{itemize}[label=\textbullet]
    \item \(\mathrm{S}, \mathrm{C}, \mathrm{M}\): Number of subjects, segmentation classes, and test-time MC dropout runs per subject, respectively.
    \item \(\mathcal{V}_s\): The set of 3D voxel coordinates for subject \(s\). All co-registered input images share the same spatial grid (\(\lvert \mathcal{V}_s\rvert = H \times W \times L\)).
   \item \( I_{i,s} \): The \(i\)-th input image for subject \(s\).
   \item \( I_{\cdot,s} \): All input images for subject \(s\), stacked as a multi-channel input.
   \item \( f^{(m)}_\theta \): Segmentation model with dropout applied during the \(m\)-th MC run.
   \item \( f^{(m)}_\theta(I_{\cdot,s})[c, v] \): The predicted probability that voxel \(v\) belongs to class \(c\).
   \item \( \nabla_{I_{i,s}} \): Gradient operator with respect to input image \(I_{i,s}\).
\end{itemize}

The OIS quantifies model dependence by measuring how sensitively the output responds to changes in each input image. We compute the gradient by first summing predicted probabilities over all voxels for a given class, then backpropagating this scalar with respect to each input. While gradients are computed voxel-wise, we aggregate them across voxels to obtain a single scalar importance score per input image. We use absolute gradient values, focusing on how strongly the image affects the output rather than direction of change. Aggregating across segmentation classes, subjects, and MC dropout runs, OIS identifies inputs that are broadly informative, robust to inter-subject variability, and stable under different model configurations. Higher OIS values indicate stronger influence on segmentation, while lower values suggest minimal relevance.\\
\textbf{Stage 2: Input modality comparison.} \\
Building on Stage 1 results, we now compare the selected multi-TI subsets against structural sequences and quantitative maps to determine the most effective input modality. The candidate multi-TI subsets identified by OIS ranking are combined with MPRAGE, FGATIR, $\mathrm{T_1}$ maps, and PD maps to form multiple candidate input configurations.
For each configuration, we train a separate 3D U-Net~\cite{cciccek20163d} on volumetric data from scratch using Dice loss and the same training protocol to ensure fair comparison. Configuration details are provided in Section~\ref{sec:input_comparison}.

\section{Experiments and Results}
\subsection{Experimental Setup}
All models in both Stage 1 and Stage 2 share the same training protocol.
The segmentation model was trained with LeakyReLU activation, instance normalization, and Adam optimization (lr=0.001, weight decay=0.0001). 
The learning rate decreases by 10\% after 5 stagnant epochs, and early stopping occurs after 15 epochs without validation improvement.  
Data augmentation including left-right flipping, affine transformations (scaling, rotation, translation), and cropping to 96×96×96 are applied to boost robustness. 
We perform 8-fold cross-validation on 24 subjects, splitting data into training, validation, and testing with a 19:2:3 ratio per fold, ensuring each subject is tested once. 
Since the ground truth labels are sparse, we use true positive rate (TPR) rather than Dice score as our evaluation metric to avoid unfairly penalizing correctly segmented but unlabeled voxels. 

\begin{figure}[!t]
\centering
\includegraphics[width = 0.95\textwidth]{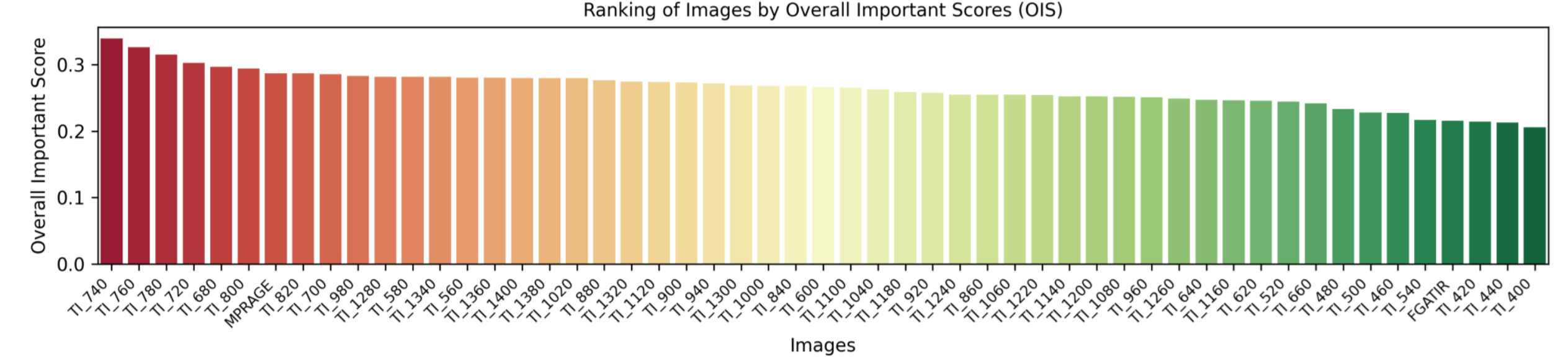}
\caption{Ranking of images by Overall Importance Score (OIS). The x-axis represents input images, while the y-axis shows their corresponding OIS values.}
\label{fig:s3-overall_importance_scores}
\end{figure}

\subsection{Selecting the Most Informative Multi-TI Images}\label{sec:3.2}
As detailed in Section~\ref{subsec:Identify the Optimal Inputs}, to identify the most informative multi-TI images, we performed gradient-based saliency analysis with test-time Monte Carlo (MC) Dropout using a 0.1 dropout rate. In 8-fold cross-validation, 24 subjects each served as the test set once and underwent 100 forward-backward passes. We then computed the Overall Importance Score (OIS) to quantify each input image’s contribution. Figure~\ref{fig:s3-overall_importance_scores} ranks the images by their OIS, where high scores indicate strong and stable contributions. Based on this ranking, we selected three candidate configurations for further evaluation: 1) the top TI at 740 ms, 2) the top two TIs at 740 and 760 ms, and 3) the top four TIs at 740, 760, 780, and 720 ms. 

\subsection{Comparative Study of Input Configurations}\label{sec:input_comparison}
Building on the selected multi-TI subsets from Section~\ref{sec:3.2}, we now compare them against structural sequences and quantitative maps across nine configurations: $\mathrm{Config_1}$: MPRAGE, $\mathrm{Config_2}$: FGATIR, $\mathrm{Config_3}$ MPRAGE+FGATIR, $\mathrm{Config_4}$: PD+$\mathrm{T_1}$ maps, $\mathrm{Config_5}$: $\mathrm{T_1}$ map, $\mathrm{Config_6}$: 51 multi-TI images (400–1400 ms, in 20 ms steps), $\mathrm{Config_7}$: the top multi-TI image at 720ms, 
$\mathrm{Config_8}$: the top two at 740 and 760 ms, 
$\mathrm{Config_9}$: the top four at 720, 740, 760, and 780 ms.

\begin{figure}[!ht]
\centering
\includegraphics[width = 1\textwidth]{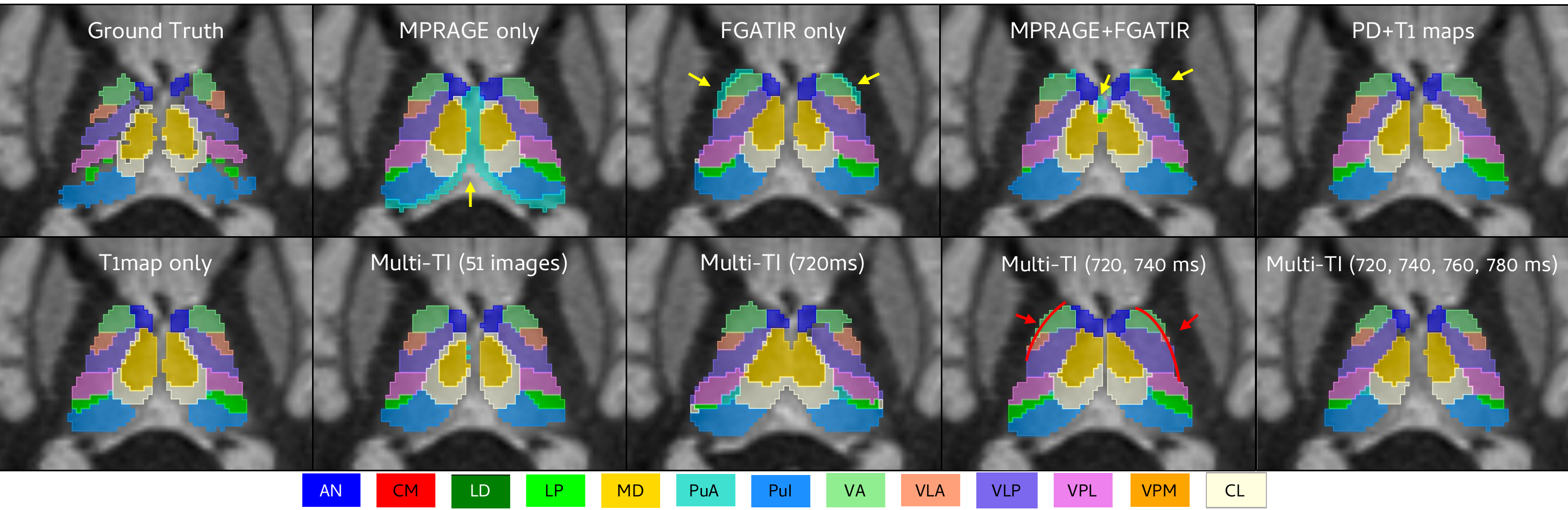}
\caption{Qualitative comparison of different input configurations.}
\label{fig:s3-ablation_results}
\end{figure}

We evaluate these configurations both quantitatively and qualitatively. Quantitatively, we computed the TPR for each thalamic nucleus and the volume-weighted average (VWA) across 13 nuclei. The mean and standard deviation (SD) of both per-nucleus TPR and VWA across 24 subjects are shown in Table~\ref{tab:s3-ablation_study}.
$\mathrm{Config_8}$ (top2 multi-TIs) achieves the highest VWA of 80.58±3.80\%, indicating that carefully selected multi-TI subsets capture essential features more effectively than MPRAGE and FGATIR, or even the full multi-TI set. $\mathrm{Config_5}$ ($\mathrm{T_1}$map) (highlighted in blue) performs comparably well, suggesting that $\mathrm{T_1}$ maps alone are sufficient for accurate segmentation. Qualitatively, as shown in Figure~\ref{fig:s3-ablation_results}, MPRAGE and FGATIR-based setups often misplace the small PuA nucleus (1.3\% of thalamus), marked by the yellow arrows. While $\mathrm{Config_8}$ (top2 multi-TIs) yields the best TPR overall, it slightly over-segments the thalamic border, marked by the red arrows. $\mathrm{Config_4}$ (PD+$\mathrm{T_1}$ maps) and $\mathrm{Config_5}$ ($\mathrm{T_1}$map) offer high-quality segmentations with minimal difference, implying PD maps contribute little beyond $\mathrm{T_1}$ maps.

For statistical validation, we use a Wilcoxon signed-rank test to compare each configuration with $\mathrm{Config_5}$ ($\mathrm{T_1}$map) on the per-nucleus TPR and VWA. A Holm–Bonferroni correction was used to control for multiple comparisons, with significance set at $p<0.05$. In Table~\ref{tab:s3-ablation_study}, \textcolor{red}{\bm{$\uparrow\uparrow$}} and \textcolor{blue}{\bm{$\downarrow\downarrow$}} indicate significantly higher or lower TPR than $\mathrm{Config_5}$ ($\mathrm{T_1}$map). Notably, there is no significant difference between $\mathrm{Config_8}$ (top2 multi-TIs) and $\mathrm{Config_5}$ ($\mathrm{T_1}$map), reinforcing that $\mathrm{T_1}$ map alone suffices. $\mathrm{Config_5}$ ($\mathrm{T_1}$map) performs similarly to $\mathrm{Config_4}$ (PD+$\mathrm{T_1}$ maps), even surpassing it in VPM, confirming that PD maps provide minimal additional benefit. Moreover, $\mathrm{Config_5}$ ($\mathrm{T_1}$map) outperforms most other experiments across multiple nuclei and VWA, establishing it as the most effective and efficient input.

\begin{table}[!t]
\centering
\fontsize{8}{8.6}\selectfont
\renewcommand{\arraystretch}{0.9}
\setlength{\tabcolsep}{0pt}
\caption{Quantitative comparison of input configurations. Mean $\pm$ SD of TPR per nucleus and the volume-weighted average (VWA) across 24 subjects, reported in percentages. Fractions under each nucleus indicates its proportional size in the thalamus. $\myuparrow$ denotes values significantly higher than $\mathrm{Config_5}$ ($\mathrm{T_1}$map), $\mydownarrow$ indicates significantly lower, and no arrow indicates no significant difference.}
\label{tab:s3-ablation_study}
\begin{tabular}{|l|c|c|c|c|c|c|c|c|c|c|c|c|c|c|}
\hline
& \begin{tabular}[c]{@{}c@{}}AN\\3.9\%\end{tabular}
& \begin{tabular}[c]{@{}c@{}}CM\\3.1\%\end{tabular}
& \begin{tabular}[c]{@{}c@{}}LD\\1.5\%\end{tabular}
& \begin{tabular}[c]{@{}c@{}}LP\\5.5\%\end{tabular}
& \begin{tabular}[c]{@{}c@{}}MD\\13.3\%\end{tabular}
& \begin{tabular}[c]{@{}c@{}}PuA\\1.3\%\end{tabular}
& \begin{tabular}[c]{@{}c@{}}Pul\\27.6\%\end{tabular}
& \begin{tabular}[c]{@{}c@{}}VA\\8.1\%\end{tabular}
& \begin{tabular}[c]{@{}c@{}}VLA\\3.4\%\end{tabular}
& \begin{tabular}[c]{@{}c@{}}VLP\\16.2\%\end{tabular}
& \begin{tabular}[c]{@{}c@{}}VPL\\6.3\%\end{tabular}
& \begin{tabular}[c]{@{}c@{}}VPM\\1.8\%\end{tabular}
& \begin{tabular}[c]{@{}c@{}}CL\\7.4\%\end{tabular}
& \begin{tabular}[c]{@{}c@{}}VWA\end{tabular} \\
\hline
MPR& 
\begin{tabular}[c]{@{}c@{}}72.47\\$\pm$\\9.71\end{tabular} &
\begin{tabular}[c]{@{}c@{}}79.70\\$\pm$\\5.35\end{tabular} &
\begin{tabular}[c]{@{}c@{}}64.50\\$\pm$\\17.38\end{tabular} &
\begin{tabular}[c]{@{}c@{}}76.72\\$\pm$\\6.95\end{tabular} &
\begin{tabular}[c]{@{}c@{}}85.52\\$\pm$\\3.91\end{tabular} &
\begin{tabular}[c]{@{}c@{}}0.00\\$\pm \mydownarrow$\\0.00\end{tabular} &
\begin{tabular}[c]{@{}c@{}}85.35\\$\pm$\\4.47\end{tabular} &
\begin{tabular}[c]{@{}c@{}}79.80\\$\pm$\\6.26\end{tabular} &
\begin{tabular}[c]{@{}c@{}}75.57\\$\pm$\\9.76\end{tabular} &
\begin{tabular}[c]{@{}c@{}}83.72\\$\pm$\\4.08\end{tabular} &
\begin{tabular}[c]{@{}c@{}}80.40\\$\pm$\\7.49\end{tabular} &
\begin{tabular}[c]{@{}c@{}}67.23\\$\pm$\\8.46\end{tabular} &
\begin{tabular}[c]{@{}c@{}}\textbf{80.40}\\$\pm\myuparrow$\\\textbf{5.84}\end{tabular} &
\begin{tabular}[c]{@{}c@{}}79.07\\$\pm$\\2.64\end{tabular} \\
\hline
FGT& 
\begin{tabular}[c]{@{}c@{}}73.21\\$\pm$\\11.62\end{tabular} &
\begin{tabular}[c]{@{}c@{}}77.67\\$\pm\mydownarrow$\\6.22\end{tabular} &
\begin{tabular}[c]{@{}c@{}}63.06\\$\pm$\\16.27\end{tabular} &
\begin{tabular}[c]{@{}c@{}}80.43\\$\pm$\\8.03\end{tabular} &
\begin{tabular}[c]{@{}c@{}}82.79\\$\pm\mydownarrow$\\5.11\end{tabular} &
\begin{tabular}[c]{@{}c@{}}42.46\\$\pm$\\34.47\end{tabular} &
\begin{tabular}[c]{@{}c@{}}\textbf{86.67}\\$\pm$\\\textbf{4.58}\end{tabular} &
\begin{tabular}[c]{@{}c@{}}78.32\\$\pm$\\9.47\end{tabular} &
\begin{tabular}[c]{@{}c@{}}76.33\\$\pm$\\9.67\end{tabular} &
\begin{tabular}[c]{@{}c@{}}82.99\\$\pm$\\6.71\end{tabular} &
\begin{tabular}[c]{@{}c@{}}\textbf{80.61}\\$\pm$\\\textbf{6.87}\end{tabular} &
\begin{tabular}[c]{@{}c@{}}69.54\\$\pm$\\9.73\end{tabular} &
\begin{tabular}[c]{@{}c@{}}61.26\\$\pm$\\8.03\end{tabular} &
\begin{tabular}[c]{@{}c@{}}79.71\\$\pm$\\2.94\end{tabular} \\
\hline
M+F& 
\begin{tabular}[c]{@{}c@{}}75.91\\$\pm$\\9.62\end{tabular} &
\begin{tabular}[c]{@{}c@{}}78.63\\$\pm$\\7.78\end{tabular} &
\begin{tabular}[c]{@{}c@{}}63.26\\$\pm$\\16.46\end{tabular} &
\begin{tabular}[c]{@{}c@{}}77.64\\$\pm$\\7.39\end{tabular} &
\begin{tabular}[c]{@{}c@{}}87.02\\$\pm$\\6.42\end{tabular} &
\begin{tabular}[c]{@{}c@{}}8.69\\$\pm\mydownarrow$\\23.03\end{tabular} &
\begin{tabular}[c]{@{}c@{}}85.15\\$\pm$\\5.50\end{tabular} &
\begin{tabular}[c]{@{}c@{}}81.27\\$\pm$\\6.33\end{tabular} &
\begin{tabular}[c]{@{}c@{}}75.23\\$\pm$\\7.05\end{tabular} &
\begin{tabular}[c]{@{}c@{}}82.57\\$\pm$\\4.39\end{tabular} &
\begin{tabular}[c]{@{}c@{}}78.86\\$\pm$\\6.58\end{tabular} &
\begin{tabular}[c]{@{}c@{}}23.65\\$\pm\mydownarrow$\\34.90\end{tabular} &
\begin{tabular}[c]{@{}c@{}}62.79\\$\pm$\\8.62\end{tabular} &
\begin{tabular}[c]{@{}c@{}}78.62\\$\pm\mydownarrow$\\3.36\end{tabular} \\
\hline
PD+$\mathrm{T_1}$& 
\begin{tabular}[c]{@{}c@{}}71.57\\$\pm$\\11.80\end{tabular} &
\begin{tabular}[c]{@{}c@{}}77.15\\$\pm$\\7.61\end{tabular} &
\begin{tabular}[c]{@{}c@{}}65.06\\$\pm$\\17.01\end{tabular} &
\begin{tabular}[c]{@{}c@{}}\textbf{81.06}\\$\pm$\\\textbf{6.06}\end{tabular} &
\begin{tabular}[c]{@{}c@{}}87.24\\$\pm$\\6.11\end{tabular} &
\begin{tabular}[c]{@{}c@{}}60.23\\$\pm$\\25.75\end{tabular} &
\begin{tabular}[c]{@{}c@{}}86.06\\$\pm$\\5.23\end{tabular} &
\begin{tabular}[c]{@{}c@{}}82.45\\$\pm$\\5.32\end{tabular} &
\begin{tabular}[c]{@{}c@{}}\textbf{76.84}\\$\pm$\\\textbf{11.85}\end{tabular} &
\begin{tabular}[c]{@{}c@{}}82.84\\$\pm$\\4.03\end{tabular} &
\begin{tabular}[c]{@{}c@{}}79.37\\$\pm$\\5.53\end{tabular} &
\begin{tabular}[c]{@{}c@{}}58.61\\$\pm\mydownarrow$\\24.95\end{tabular} &
\begin{tabular}[c]{@{}c@{}}61.12\\$\pm$\\7.29\end{tabular} &
\begin{tabular}[c]{@{}c@{}}80.34\\$\pm$\\2.89\end{tabular} \\
\hline
\rowcolor{blue!10}
$\mathrm{T_1}$map& 
\begin{tabular}[c]{@{}c@{}}76.44\\$\pm$\\10.09\end{tabular} &
\begin{tabular}[c]{@{}c@{}}\textbf{81.68}\\$\pm$\\\textbf{6.91}\end{tabular} &
\begin{tabular}[c]{@{}c@{}}63.63\\$\pm$\\17.65\end{tabular} &
\begin{tabular}[c]{@{}c@{}}78.90\\$\pm$\\5.94\end{tabular} &
\begin{tabular}[c]{@{}c@{}}87.76\\$\pm$\\5.82\end{tabular} &
\begin{tabular}[c]{@{}c@{}}40.65\\$\pm$\\32.84\end{tabular} &
\begin{tabular}[c]{@{}c@{}}85.35\\$\pm$\\4.61\end{tabular} &
\begin{tabular}[c]{@{}c@{}}76.27\\$\pm$\\13.15\end{tabular} &
\begin{tabular}[c]{@{}c@{}}76.69\\$\pm$\\7.22\end{tabular} &
\begin{tabular}[c]{@{}c@{}}82.79\\$\pm$\\4.54\end{tabular} &
\begin{tabular}[c]{@{}c@{}}80.08\\$\pm$\\5.72\end{tabular} &
\begin{tabular}[c]{@{}c@{}}\textbf{71.15}\\$\pm$\\\textbf{7.68}\end{tabular} &
\begin{tabular}[c]{@{}c@{}}64.66\\$\pm$\\6.30\end{tabular} &
\begin{tabular}[c]{@{}c@{}}80.20\\$\pm$\\2.86\end{tabular} \\
\hline
TI(51) & 
\begin{tabular}[c]{@{}c@{}}73.79\\$\pm$\\10.48\end{tabular} &
\begin{tabular}[c]{@{}c@{}}73.18\\$\pm$\\13.32\end{tabular} &
\begin{tabular}[c]{@{}c@{}}48.31\\$\pm$\\23.51\end{tabular} &
\begin{tabular}[c]{@{}c@{}}76.73\\$\pm$\\8.08\end{tabular} &
\begin{tabular}[c]{@{}c@{}}85.06\\$\pm$\\5.53\end{tabular} &
\begin{tabular}[c]{@{}c@{}}23.32\\$\pm$\\31.76\end{tabular} &
\begin{tabular}[c]{@{}c@{}}86.25\\$\pm$\\4.44\end{tabular} &
\begin{tabular}[c]{@{}c@{}}80.76\\$\pm$\\5.16\end{tabular} &
\begin{tabular}[c]{@{}c@{}}75.79\\$\pm$\\8.72\end{tabular} &
\begin{tabular}[c]{@{}c@{}}82.89\\$\pm$\\3.25\end{tabular} &
\begin{tabular}[c]{@{}c@{}}80.33\\$\pm$\\5.52\end{tabular} &
\begin{tabular}[c]{@{}c@{}}68.58\\$\pm$\\8.85\end{tabular} &
\begin{tabular}[c]{@{}c@{}}67.29\\$\pm$\\7.19\end{tabular} &
\begin{tabular}[c]{@{}c@{}}79.66\\$\pm$\\2.70\end{tabular} \\
\hline
TI(top1)& 
\begin{tabular}[c]{@{}c@{}}53.12\\$\pm\mydownarrow$\\18.09\end{tabular} &
\begin{tabular}[c]{@{}c@{}}73.96\\$\pm\mydownarrow$\\14.41\end{tabular} &
\begin{tabular}[c]{@{}c@{}}\textbf{66.33}\\$\pm$\\\textbf{20.37}\end{tabular} &
\begin{tabular}[c]{@{}c@{}}78.67\\$\pm$\\8.00\end{tabular} &
\begin{tabular}[c]{@{}c@{}}54.81\\$\pm\mydownarrow$\\17.21\end{tabular} &
\begin{tabular}[c]{@{}c@{}}\textbf{68.70}\\$\pm\myuparrow$\\\textbf{9.28}\end{tabular} &
\begin{tabular}[c]{@{}c@{}}83.78\\$\pm$\\6.52\end{tabular} &
\begin{tabular}[c]{@{}c@{}}52.84\\$\pm\mydownarrow$\\17.26\end{tabular} &
\begin{tabular}[c]{@{}c@{}}74.21\\$\pm$\\10.75\end{tabular} &
\begin{tabular}[c]{@{}c@{}}77.98\\$\pm$\\14.06\end{tabular} &
\begin{tabular}[c]{@{}c@{}}73.91\\$\pm$\\14.21\end{tabular} &
\begin{tabular}[c]{@{}c@{}}68.63\\$\pm$\\8.74\end{tabular} &
\begin{tabular}[c]{@{}c@{}}59.04\\$\pm$\\14.03\end{tabular} &
\begin{tabular}[c]{@{}c@{}}70.90\\$\pm\mydownarrow$\\4.22\end{tabular} \\
\hline
TI(top2)& 
\begin{tabular}[c]{@{}c@{}}\textbf{77.09}\\$\pm$\\\textbf{10.12}\end{tabular} &
\begin{tabular}[c]{@{}c@{}}79.73\\$\pm$\\8.25\end{tabular} &
\begin{tabular}[c]{@{}c@{}}58.08\\$\pm$\\17.60\end{tabular} &
\begin{tabular}[c]{@{}c@{}}77.75\\$\pm$\\8.20\end{tabular} &
\begin{tabular}[c]{@{}c@{}}\textbf{89.84}\\$\pm$\\\textbf{4.21}\end{tabular} &
\begin{tabular}[c]{@{}c@{}}66.83\\$\pm$\\10.66\end{tabular} &
\begin{tabular}[c]{@{}c@{}}84.04\\$\pm$\\6.13\end{tabular} &
\begin{tabular}[c]{@{}c@{}}\textbf{82.98}\\$\pm$\\\textbf{4.78}\end{tabular} &
\begin{tabular}[c]{@{}c@{}}72.23\\$\pm$\\11.18\end{tabular} &
\begin{tabular}[c]{@{}c@{}}81.73\\$\pm$\\4.61\end{tabular} &
\begin{tabular}[c]{@{}c@{}}78.18\\$\pm$\\5.99\end{tabular} &
\begin{tabular}[c]{@{}c@{}}67.92\\$\pm$\\8.83\end{tabular} &
\begin{tabular}[c]{@{}c@{}}67.83\\$\pm$\\8.17\end{tabular} &
\begin{tabular}[c]{@{}c@{}}\textbf{80.58}\\$\pm$\\\textbf{3.80}\end{tabular} \\
\hline
TI(top4)& 
\begin{tabular}[c]{@{}c@{}}74.06\\$\pm$\\13.16\end{tabular} &
\begin{tabular}[c]{@{}c@{}}54.83\\$\pm\mydownarrow$\\19.48\end{tabular} &
\begin{tabular}[c]{@{}c@{}}65.79\\$\pm$\\22.49\end{tabular} &
\begin{tabular}[c]{@{}c@{}}78.52\\$\pm$\\9.99\end{tabular} &
\begin{tabular}[c]{@{}c@{}}84.84\\$\pm$\\6.35\end{tabular} &
\begin{tabular}[c]{@{}c@{}}49.81\\$\pm$\\25.98\end{tabular} &
\begin{tabular}[c]{@{}c@{}}85.22\\$\pm$\\5.29\end{tabular} &
\begin{tabular}[c]{@{}c@{}}81.82\\$\pm$\\5.69\end{tabular} &
\begin{tabular}[c]{@{}c@{}}72.44\\$\pm$\\8.78\end{tabular} &
\begin{tabular}[c]{@{}c@{}}\textbf{84.13}\\$\pm$\\\textbf{4.87}\end{tabular} &
\begin{tabular}[c]{@{}c@{}}77.61\\$\pm$\\6.84\end{tabular} &
\begin{tabular}[c]{@{}c@{}}68.58\\$\pm$\\11.14\end{tabular} &
\begin{tabular}[c]{@{}c@{}}69.05\\$\pm$\\6.40\end{tabular} &
\begin{tabular}[c]{@{}c@{}}78.87\\$\pm$\\3.23\end{tabular} \\
\hline
\end{tabular}
\end{table}

\section{Discussions and Conclusions}
Our experiments show that the essential tissue contrast captured by multi-TI images is inherently encoded in the $\mathrm{T_1}$ map. Since multi-TI images are derived from the $\mathrm{T_1}$ map, direct segmentation using the $\mathrm{T_1}$ map reduces preprocessing demands while preserving key information. Additionally, multi-TI images must be carefully selected, as blindly incorporating all 51 multi-TI images does not outperform the $\mathrm{T_1}$map-only approach, suggesting that an unfiltered multi-TI strategy is less effective than simply leveraging the $\mathrm{T_1}$ map itself. Moreover, PD maps offer no added value beyond $\mathrm{T_1}$ maps, reinforcing $\mathrm{T_1}$ maps as the primary source of contrast. Although some $\mathrm{T_1}$-map features are not visually apparent, computational models can extract meaningful structures beyond human perception. Given this, is there still value in computing multi-TI images? We argue yes, as they offer enhanced visual contrast that supports manual delineation.

A practical concern is the availability of high-quality $\mathrm{T_1}$ maps. When not directly accessible, they can be estimated from MPRAGE or FGATIR, acquired via MP2RAGE~\cite{marques2010mp2rage}, or synthesized from MPRAGE using learning-based methods~\cite{hays2024revisiting}. 
Our approach has several limitations: 
First, the dataset size is relatively small, though we mitigate this with 8-fold cross-validation and statistical evaluation. 
Second, external validation remains challenging because most public datasets (e.g., ADNI, HCP, BLSA, IXI) lack key contrasts such as FGATIR or $\mathrm{T_1}$ maps, limiting their direct use for evaluation. Future work will focus on synthesizing $\mathrm{T_1}$ maps in these datasets and training more generalizable models. For datasets without manual labels, we plan to explore semi-supervised or pseudo-labeling strategies.
Third, we do not explore all imaging modalities; our comparisons focus mainly on T1w-based inputs. However, modalities such as T2w, diffusion MRI, and structural connectivity are excluded intentionally. T2w images are rarely used for thalamic segmentation due to limited intra-thalamic contrast. Diffusion and connectivity features capture microstructural information that may not structurally align with T1w anatomy. Since our manual labels are defined on T1w images, evaluating non-T1w modalities against them would inherently disadvantage these inputs and result in unfair comparisons.

In conclusion, in this paper, we systematically evaluated structural sequences (MPRAGE, FGATIR), quantitative maps (PD, $\mathrm{T_1}$ maps), and multi-TI images for thalamic nuclei segmentation. To select the most informative multi-TI images, we introduced a novel approach that integrates gradient‐based saliency analysis with test‐time Monte Carlo dropout and proposed the Overall Importance Score to quantify each image’s contribution. This evaluation framework is generalizable to any input selection or feature attribution task. Moreover, our findings show that $\mathrm{T_1}$map-based segmentation offers the best balance of accuracy and efficiency, making it the most practical choice for thalamic nuclei segmentation.

\section*{Acknowledgment and Disclaimer}
This research was supported in part by the Intramural Research Program of the National Institutes of Health (NIH). The contributions of the NIH authors were made as part of their official duties as NIH federal employees, are in compliance with agency policy requirements, and are considered Works of the United States Government. However, the findings presented in this paper are those of the authors and do not necessarily reflect the views of the NIH or the U.S. Department of Health and Human Services.
The work is also supported in part by NIH through the National Institute of Neurological Disorders and Stroke grant R01-NS105503~(PIs: Zhuo \& Prince).

\bibliographystyle{splncs04}
\bibliography{reference}

\end{document}